\documentclass{ws-ijmpcs}

\begin{document}

\markboth{C. Dilks}
{Forward Neutral Pion Double Helicity Asymmetry at STAR}

%
\catchline{}{}{}{}{}
%

\title{
  Double Helicity Asymmetries of Forward Neutral Pions
  from $\sqrt{s}=510$ GeV $pp$ Collisions at STAR
}

\author{Christopher J. Dilks}

\address{Department of Physics, Pennsylvania State University\\
University Park, PA 16802,
United States\\
cjd5150@psu.edu}

\maketitle

\begin{history}
\received{Day Month Year}
\revised{Day Month Year}
\published{Day Month Year}
\end{history}

\begin{abstract}
Longitudinally polarized $p+p$ scattering experiments provide access to 
gluon polarization via measurement of the double helicity asymmetry, $A_{LL}$.
At the completion of the 2013 RHIC running period, a significant dataset of $\pi^{0}$s corresponding to an integrated luminosity of 46 $\text{pb}^{-1}$ (2012) and 8 $\text{pb}^{-1}$ (2013)
produced from polarized $p+p$ scattering at $\sqrt{s}=510$ GeV with an average
beam polarization of approximately $50\%$ was acquired. The $\pi^{0}$ kinematics 
were measured via isolation cones by the STAR Forward Meson Spectrometer, an electromagnetic calorimeter 
covering a forward pseudorapidity range of $2.6 < \eta <4$.
The asymmetric $qg \to qg$ subprocess becomes more dominant in this forward region 
than in the midrapidity region; furthermore, asymmetry measurements in the forward region are sensitive to low-$x$ gluons. Progress on $A_{LL}$ determined from 
forward $\pi^{0}$ events, complementing previous midrapidity measurements, are presented.
\keywords{double helicity asymmetry; longitudinal double spin asymmetry; forward $\pi^0$}
\end{abstract}

\ccode{PACS numbers:}

\section{Introduction}
The spin of the proton is decomposed in terms of the constituent quark and gluon 
spins and angular momenta. The contribution from the quark spin has been measured to account
for approximately 30\% of the proton spin.\cite{phenix_A_LL} While the quark 
and gluon angular momenta have not been measured, the gluon helicity contribution is coming 
into focus. The gluon helicity contribution is described by the zeroth moment of the 
polarized gluon parton distribution function (PDF), $\Delta g(x)$, defined as the difference in the 
unpolarized gluon PDF for which gluon helicity is aligned with the proton helicity and that for which
the gluon and proton helicities are anti-aligned. According to the most recent global analyses, the 
$\Delta g(x)$ distribution is positive for $x>0.05$, but remains largely unconstrained for 
$x<0.05$.\cite{dssv,nnpdf}

In $pp$ scattering, the double helicity asymmetry for $\pi^0$ production, $A_{LL}^{\pi^0}$, is defined 
as the ratio of the polarized cross section to the unpolarized cross section.
Assuming factorization validity for the $pp\to\pi^0X$ process, $A_{LL}^{\pi^0}$ may be expressed as
\begin{equation}
\label{ALL_def}
A_{LL}^{\pi^0}=
\frac{\Delta\sigma\left(pp\to\pi^0X\right)}{\sigma\left(pp\to\pi^0X\right)}=
\frac{\sum_{abc}{\Delta f_a\otimes\Delta f_b\otimes\Delta\hat{\sigma}\left(ab\to cX\right)
\otimes D_c^{\pi^0}}}
{\sum_{abc}{f_a\otimes f_b\otimes\hat{\sigma}\left(ab\to cX\right)\otimes D_c^{\pi^0}}},
\end{equation}
where the summations run over partons, $\Delta f_i$ and $f_i$ are polarized and unpolarized PDFs, $\Delta\hat{\sigma}$ and $\hat{\sigma}$ are parton-level polarized and unpolarized cross sections, and $D_c^{\pi^0}$ is the fragmentation function for $\pi^0$ hadronization.
Of the probability density distributions which enter equation \eqref{ALL_def}, the least constrained quantity is $\Delta g(x)$, which in turn makes $A_{LL}$ an observable sensitive to gluon polarization.\cite{jet_production,star_pi0_A_LL} 

The least-understood $x<0.05$ behavior of $\Delta g(x)$ is kinematically accessible by a measurement of $A_{LL}$ for hadroproduction at forward pseudorapidities.\cite{forward_pi0_low_x} Forward jets are typically produced by subprocesses with a low-$x$ parton scattering on a much higher-$x$ parton. Since the $qg \to qg$ subprocess becomes more dominant in the forward region and the gluon PDF dominates at low-$x$, foward jets are sensitive to low-$x$ gluon phenomena.

\section{Forward Calorimetry at STAR}
\begin{figure}[pt]
\centerline{\includegraphics[width=8.0cm]{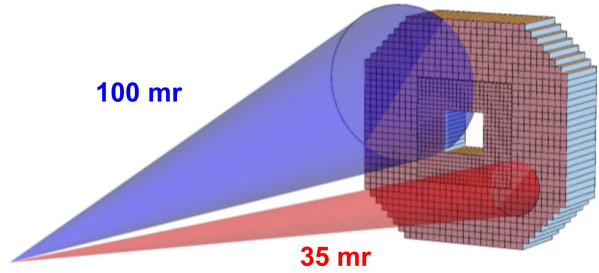}}
\vspace*{8pt}
\caption{Schematic of the FMS lead glass cells, shown with relative isolation cone sizes of 35 mrad (red) and 100 mrad (blue). \label{cones}}
\end{figure}

The Forward Meson Spectrometer (FMS) is an electromagnetic calorimeter in the STAR experimental hall of RHIC.
Covering full azimuth and a forward pseudorapidity range of $2.6 < \eta < 4$, the FMS is composed of an array of lead glass cells, where each cell is backed by a photomultiplier tube; see Fig.~\ref{cones} for a schematic layout of the cells.
The primary observable is a pair of photon clusters, which originate from the primary decay mode of $\pi^0$s.
Measuring the photon energies $E_1$ and $E_2$ along with the cluster separation allows for the determination of the invariant mass and the subsequent $\pi^0$ event reconstruction. 
Each candidate $\pi^0$ event is characterized by an invariant mass $M_{\gamma\gamma}$, di-photon energy $E_{\gamma\gamma}=E_1+E_2$, and transverse momentum $p_T$.

\begin{figure}[pt]
\centerline{\includegraphics[width=11.0cm]{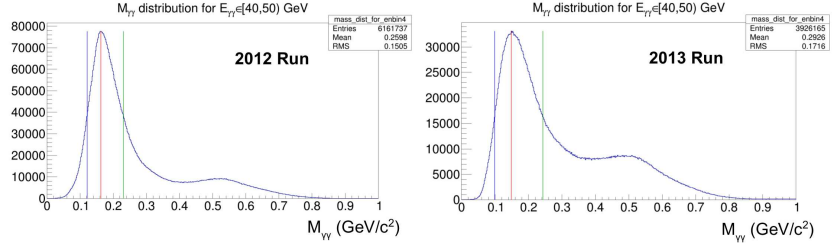}}
\vspace*{8pt}
\caption{$M_{\gamma\gamma}$ distribution for $40 \leq E_{\gamma\gamma} < 50$. Red line indicates $\pi^0$ mass; blue and green lines denote lower and upper bounds on the $M_{\gamma\gamma}$ range analyzed for this $E_{\gamma\gamma}$ range.  \label{mass_dists}}
\end{figure}

The data set used in this analysis was obtained during the 2012 and 2013 RHIC running periods from longitudinally polarized $pp$ collisions with center of mass energy $\sqrt{s}=510$ GeV. Altogether, the FMS acquired an integrated luminosity of 46 $\text{pb}^{-1}$ in 2012 and 8 $\text{pb}^{-1}$ in 2013. The $\pi^0$ events were selected based on the following criteria:
\begin{itemlist}
  \item $p_T \geq 2.5$ GeV/c for 2012 data and $p_T \geq 2.0$ GeV/c for 2013 data,
  \item $p_T < 10$ GeV/c,
  \item $30 \leq E_{\gamma\gamma} < 100$ GeV,
  \item $\left|E_1-E_2\right|/E_{\gamma\gamma}<0.8$,
  \item $E_{\gamma\gamma}$-dependent $M_{\gamma\gamma}$ cut,
  \item isolation cone cut (both 35 mrad and 100 mrad analyzed).
\end{itemlist}
The $E_{\gamma\gamma}$-dependent $M_{\gamma\gamma}$ cut was used to account for the overall increase in $M_{\gamma\gamma}$ reconstruction with increasing $E_{\gamma\gamma}$; this effect serves as a guide for and can be fixed by FMS calibration improvement. An example $M_{\gamma\gamma}$ distribution for $40 \leq E_{\gamma\gamma} < 50$ GeV is shown in Fig.~\ref{mass_dists}.

The data were analyzed using two $\pi^0$ isolation cone sizes of 35 mrad and 100 mrad, shown in Fig.~\ref{cones}. The use of isolation cones was motivated by the dependence of the transverse single-spin asymmetry $A_N$ on $\pi^0$ isolation: more isolated $\pi^0$s exhibited higher asymmetries.\cite{pi0_isolation} A goal of this study was to verify that $A_{LL}$ is not dependent on $\pi^0$ isolation.

\section{Relative Luminosity and $A_{LL}^{\pi^0}$ Systematic Uncertainty}
Experimentally, it is easiest to use $\pi^0$ counts to measure $A_{LL}^{\pi^0}$ rather than cross sections.
Letting $h_1$ and $h_2$ denote the initial proton helicity signs, the proton spin-dependent cross sections $\sigma_{h_1h_2}$ for $\pi^0$ production may be re-expressed in terms of the $\pi^0$ yield counts $N_{h_1h_2}$ and integrated luminosities $L_{h_1h_2}$. 
With the relative luminosity $R_3$, defined as 
\begin{equation}
\label{rellum}
R_3=\frac{L_{++}+L_{--}}{L_{+-}+L_{-+}},
\end{equation}
$A_{LL}^{\pi^0}$ may rewritten in terms of the yields and proton beam polarizations $P_a$ and $P_b$ as
\begin{equation}
\label{ALL}
A_{LL}^{\pi^0}=\frac{1}{P_aP_b}
\frac{\left(N_{++}+N_{--}\right)-R_3\left(N_{+-}+N_{-+}\right)}
{\left(N_{++}+N_{--}\right)+R_3\left(N_{+-}+N_{-+}\right)}.
\end{equation}
The beam polarizations were measured independently by the RHIC polarimetry group, and were typically $\sim 50\%$. 

\begin{figure}[pt]
\centerline{\includegraphics[width=13.0cm]{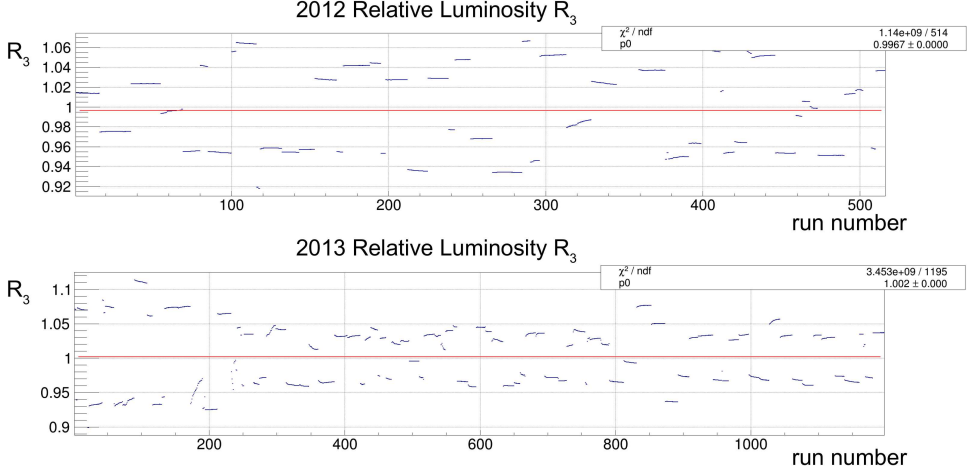}}
\vspace*{8pt}
\caption{Relative luminosity $R_3$ value for each data-taking run, labeled by ``run number.''\label{rellum_data}}
\end{figure}

The relative luminosity was measured using the Vertex Position Detector (VPD) and cross-checked with the Zero Degree Calorimeter (ZDC).
Both the VPD and ZDC are called ``binary counting detectors'': for each proton bunch crossing at STAR, they registered whether or not a particle passed through them with enough energy to satisfy a threshold condition, without any other information recorded. The VPD and ZDC are both duplicated and positioned symmetrically on each side of the $pp$ interaction point. This setup makes it possible to define coincidence triggers, which demand a count on each side of the VPD or ZDC within a defined timing interval. For each data-taking run (typically a 30 minute period), the relative luminosity was computed using an average of VPD single-side triggers and VPD coincidence triggers. 
The relative luminosity for each data-taking run is shown in Fig.~\ref{rellum_data}.

\begin{figure}[pt]
\centerline{\includegraphics[width=13.0cm]{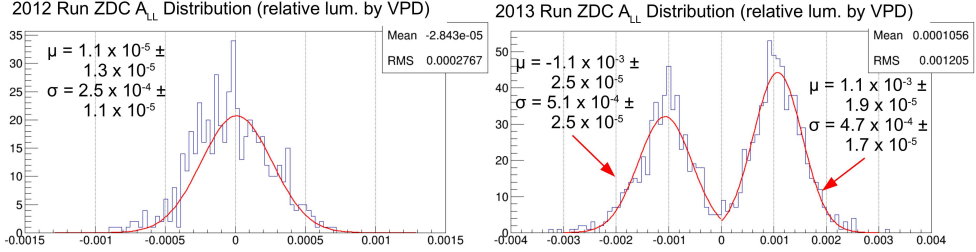}}
\vspace*{8pt}
\caption{Distribution of $A_{LL}^{ZDC/VPD}$ for 2012 and 2013 data sets; $\mu$ and $\sigma$ are the mean and standard deviation of the Gaussian fits (red line) and the overall distribution mean and RMS of the full distribution is written in the upper-right corner for both data sets\label{scaler_ALL}}
\end{figure}

The VPD and ZDC were also used to constrain a systematic uncertainty on the $A_{LL}^{\pi^0}$ measurement.
The double helicity asymmetry was also measured using ZDC single-side triggers while using the VPD coincidence triggers as a relative luminosity.
This asymmetry, denoted $A_{LL}^{ZDC/VPD}$, was also measured for each data-taking run and is shown as a distribution in Fig.~\ref{scaler_ALL}.
The distributions are fit to Gaussian functions, where $\mu$ and $\sigma$ are the distribution mean and standard deviation, respectively.
For the 2013 data set, the value of $A_{LL}^{ZDC/VPD}$ was correlated with the choice of spin pattern, where half of the spin patterns yielded negative values and the other half yielded positive values; this effect is still under investigation. The distribution means give a value of $\langle A_{LL}^{ZDC/VPD}\rangle=-2.8\times 10^{-5}$ for 2012 data and $1.1\times 10^{-4}$ for 2013 data. 

The contribution to the $A_{LL}^{\pi^0}$ systematic uncertainty from the relative luminosity measurement was conservatively estimated as the sum of the $A_{LL}^{ZDC/VPD}$ distribution mean and the standard deviation of the Gaussian fit. 
For the 2013 data set, the larger of the two standard deviations was used. 
The systematic uncertainty values obtained from 2012 and 2013 were combined to a single value by using $\pi^0$ statistics to compute a weighted average; this computation was performed for each $p_T$ and $E_{\gamma\gamma}$ bin used for the $A_{LL}^{\pi^0}$ extraction.
The systematic uncertainty contribution was $2.8\times 10^{-4}$ for the 2012 data set and $6.2\times 10^{-4}$ for the 2013 data set.

\section{Forward $A_{LL}^{\pi^0}$ Measurement}
\begin{figure}[pt]
\centerline{\includegraphics[width=10.0cm]{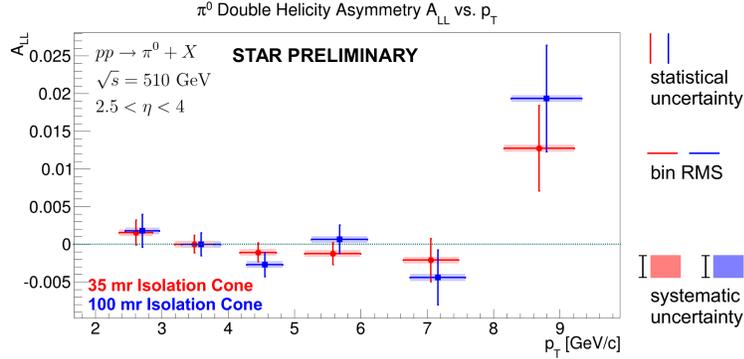}}
\vspace*{8pt}
\caption{Double helicity asymmetry $A_{LL}^{\pi^0}$ in $p_T$ bins for two $\pi^0$ isolation cones; horizontal cyan line indicates zero, vertical error bars are statistical uncertainties, and horizontal error bars indicate the RMS of each $p_T$ bin. The vertical spread of the shading indicates the systematic uncertainty contribution from the relative luminosity. The 100 mrad isolation cone data points are offset by $p_T+0.1$ GeV/c for visibility.\label{ALL_pt}}
\end{figure}

\begin{figure}[pt]
\centerline{\includegraphics[width=10.0cm]{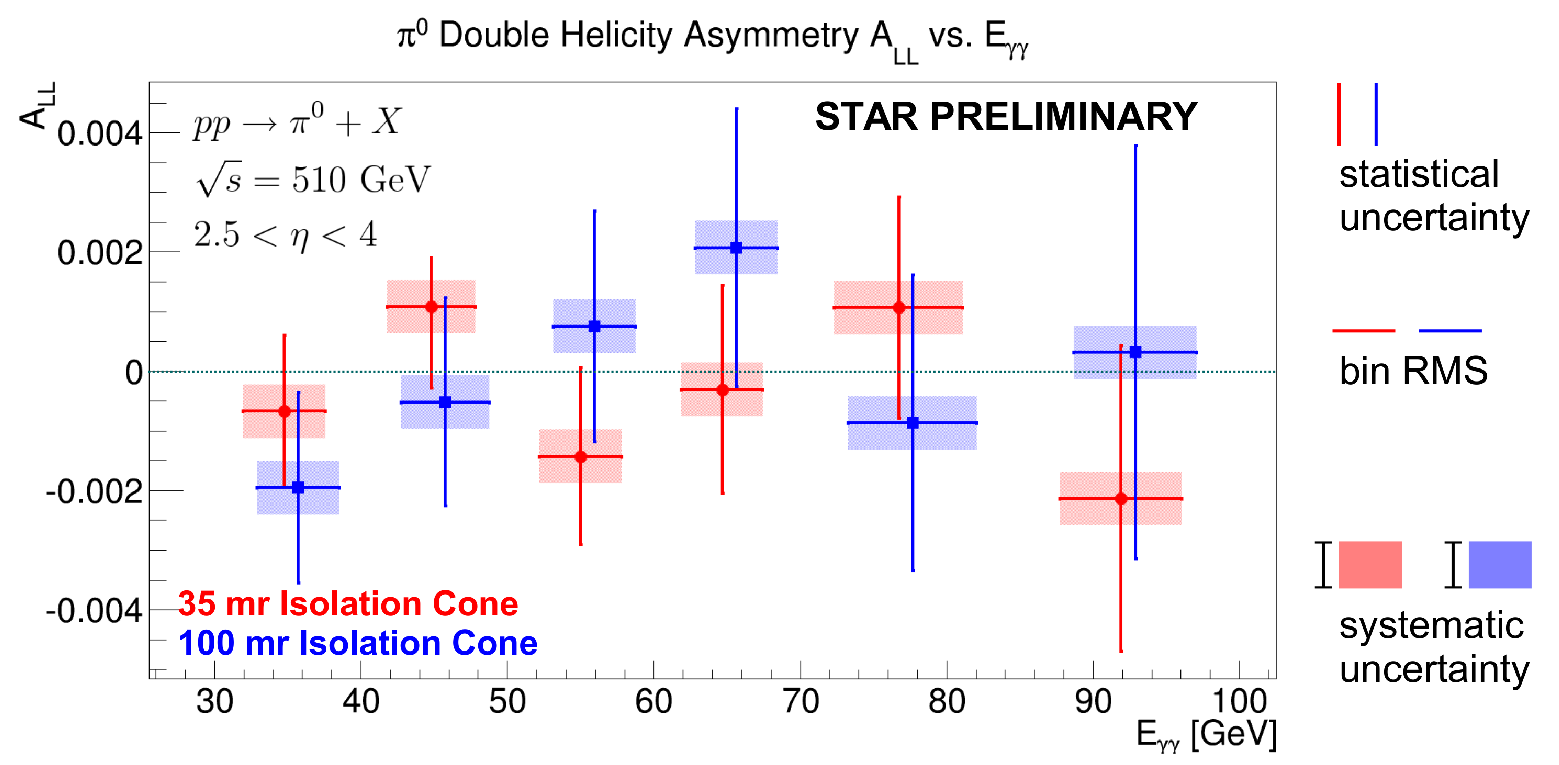}}
\vspace*{8pt}
\caption{Double helicity asymmetry $A_{LL}^{\pi^0}$ in $E_{\gamma\gamma}$ bins for two $\pi^0$ isolation cones; horizontal cyan line indicates zero, vertical error bars are statistical uncertainties, and horizontal error bars indicate the RMS of each $E_{\gamma\gamma}$ bin. The vertical spread of the shading indicates the systematic uncertainty contribution from the relative luminosity. The 100 mrad isolation cone data points are offset by $E_{\gamma\gamma}+1$ GeV for visibility.\label{ALL_en}}
\end{figure}

The measurement of $A_{LL}^{\pi^0}$ is plotted in $p_T$ bins in Fig.~\ref{ALL_pt} and in $E_{\gamma\gamma}$ bins in Fig.~\ref{ALL_en}. 
A constant fit on the 35 mrad isolation cone data returns a value of $A_{LL}^{\pi^0}=-2.5\times 10^{-4}\pm 6.5\times 10^{-4}$. For the 100 mrad isolation cone, the constant fit returns $A_{LL}^{\pi^0}=-3.3\times 10^{-4}\pm 8.4\times 10^{-4}$. 
There is no indication of nonzero $A_{LL}^{\pi^0}$ within the present statistical and systematic uncertainties. Furthermore, there is no observed $\pi^0$ isolation-dependence of $A_{LL}^{\pi^0}$.
Additional systematic uncertainty contributions, such as trigger bias and initial proton transversity remain under investigation, but are likely to be sub-dominant. 


\section{Conclusion}
A measurement of the double helicity asymmetry for forward-produced neutral pions, $A_{LL}^{\pi^0}$, has been presented. 
In particular, the forward pseudorapidity range of $2.6 < \eta < 4$ allows one to constrain $\Delta g(x)$ for $x<0.05$.
The asymmetry is consistent with zero within the statistical and systematic uncertainties.
Although this measurement was performed using $\pi^0$ isolation cones, an inclusive measurement is underway.



\bibliographystyle{ws-ijmpcs}
\bibliography{sources}

\end{document}